\documentstyle[12pt,axodraw]{article}

\catcode`@=12
\begin{document}
\thispagestyle{empty}
\begin{flushright} 
UCRHEP-T313\\ 
August 2001\
\end{flushright}
\vspace{0.5in}
\begin{center}
{\LARGE \bf Supersymmetric Model of Muon Anomalous\\ Magnetic Moment and 
Neutrino Masses\\}
\vspace{1.5in}
{\bf Rathin Adhikari$^{a,b}$, Ernest Ma$^c$, and G. Rajasekaran$^d$\\}
\vspace{0.2in}
{\sl $^a$ Physics Department, Jadavpur University, Kolkata (Calcutta) 700032, 
India\\}
\vspace{0.1in}
{\sl $^b$ Physics Department, University of Calcutta, Kolkata (Calcutta) 
700009, India\\}
\vspace{0.1in}
{\sl $^c$ Physics Department, University of California, Riverside, 
California 92521\\}
\vspace{0.1in}
{\sl $^d$ Institute of Mathematical Sciences, Chennai (Madras) 600113, India\\}
\vspace{1.5in}
\end{center}
\begin{abstract}\
We propose the novel lepton-number relationship $L_\tau = L_e + L_\mu$, which 
is uniquely realized by the interaction $(\hat \nu_e \hat \mu - \hat e \hat 
\nu_\mu) \hat \tau^c$ in supersymmetry and may account for a possibly large 
muon anomalous magnetic moment.  Neutrino masses (with bimaximal mixing) may 
be generated from the spontaneous and soft breaking of this lepton symmetry.
\end{abstract}
\newpage
\baselineskip 24pt

In the minimal standard model of particle interactions, the 3 lepton numbers 
$L_e$, $L_\mu$, $L_\tau$, are separately conserved automatically.  If it is 
extended to include supersymmetry, the assignment of $L_e$, $L_\mu$, and 
$L_\tau$ becomes more complicated.  However, it has been shown some time ago 
\cite{mn} that there are actually 17 well-defined models: 1 with 3 lepton 
numbers, i.e. the minimal supersymmetric standard model (MSSM), 6 with 2 
lepton numbers, 9 with 1 lepton number, and 1 with no lepton number, i.e. the 
general $R$-parity violating (but baryon-number conserving) supersymmetric 
model.  Three such models are particularly interesting, because they require 
only one additional term in the superpotential beyond that of the MSSM, i.e. 
\begin{equation}
\hat W = h (\hat \nu_e \hat \mu - \hat e \hat \nu_\mu) \hat \tau^c
\end{equation}
and its two obvious permutations.  These terms are unique because they are 
the only ones allowed by the conservation of two lepton numbers \cite{mn} 
with the pattern $e \sim (1,0)$, $\mu \sim (0,1)$, and $\tau \sim (1,1)$ 
for the example given above.

In this paper we will show that this extra term allows a possibly large 
contribution to the anomalous magnetic moment of the muon \cite{g-2}, 
independent of other possible MSSM contributions \cite{susyg2}.  We then 
break this symmetry softly and spontaneously, and show that neutrino masses 
(with bimaximal mixing) are easily obtained for a natural explanation of 
the atmospheric \cite{atm} and solar \cite{solar} neutrino observations. 

The interaction terms of the Lagrangian resulting from Eq.~(1) are given by
\begin{equation}
{\cal L}_{int} = h (\nu_e \mu - e \nu_\mu) \tilde \tau^c + h (\nu_e \tau^c 
\tilde \mu - e \tau^c \tilde \nu_\mu) + h (\mu \tau^c \tilde \nu_e - \nu_\mu 
\tau^c \tilde e) + h.c.
\end{equation}
Hence there are 2 contributions to the muon anomalous magnetic moment as 
shown in Fig.~1.  They are easily evaluated \cite{kkl} and we obtain 
\begin{equation}
\Delta a_\mu = {h^2 m_\mu^2 \over 96 \pi^2} \left( {2 \over m^2_{\tilde 
\nu_e}} - {1 \over m^2_{\tilde \tau^c}} \right).
\end{equation}
Similarly,
\begin{eqnarray}
\Delta a_e &=& {h^2 m_e^2 \over 96 \pi^2} \left( {2 \over m^2_{\tilde 
\nu_\mu}} - {1 \over m^2_{\tilde \tau^c}} \right), \\ 
\Delta a_\tau &=& {h^2 m_\tau^2 \over 96 \pi^2} \left( {2 \over m^2_{\tilde 
\nu_e}} +{2 \over m^2_{\tilde \nu_\mu}} - {1 \over m^2_{\tilde e}} - {1 \over 
m^2_{\tilde \mu}} \right).
\end{eqnarray}

Of all the possible effective four-fermion interactions which can be derived 
from Eq.~(2), only two are easily accessible experimentally: $\mu \to e 
\nu_\mu \bar \nu_e$ through $\tilde \tau^c$ exchange \cite{bgh} and 
$e^+ e^- \to \tau^+ \tau^-$ through $\tilde \nu_\mu$ exchange.  For 
simplicity, both $\tilde \tau^c$ and $\tilde \nu_\mu$ may be assumed to be 
heavy, say a few TeV, 
then the coupling $h$ is allowed to be of order unity in Eq.~(2).  To obtain 
$\Delta a_\mu \sim 10^{-9}$ to account for the possible discrepancy of the 
experimental value \cite{g-2} with the standard-model expectation \cite{qcd}, 
we need $\tilde \nu_e$ to be relatively light, say around 200 GeV.

Our model as it stands forbids neutrino masses because it conserves $L_e$ and 
$L_\mu$ (with $L_\tau = L_e + L_\mu$).  Consider now the \underline {soft} 
breaking of these lepton numbers by the terms
\begin{equation}
\mu_\alpha (\hat l_\alpha \hat h_2^+ - \hat \nu_\alpha \hat h_2^0)
\end{equation}
in the superpotential, i.e. the so-called bilinear $R$-parity violation 
\cite{cf}.  In that case, the $4 \times 4$ neutralino mass matrix of the 
MSSM must be expanded to include the 3 neutrinos as well to form a 
$7 \times 7$ mass matrix.  It is well-known that \underline {one} tree-level 
mass, corresponding to a linear combination of $\nu_e$, $\nu_\mu$, and 
$\nu_\tau$ is now obtained.  In this scenario, the scalar neutrinos also 
acquire nonzero vacuum expectation values \cite{drv} and one-loop radiative 
neutrino masses are possible \cite{brpv}.  To fit the present data on 
atmospheric \cite{atm} and solar \cite{solar} neutrino oscillations, 
restrictions on the parameters of the MSSM are implied.

In our model there is another, unrestricted source of radiative neutrino 
mass, as shown in Fig.~2.  This gives a contribution only to the off-diagonal 
$\nu_e \nu_\mu$ term.  Hence our effective $3 \times 3$ neutrino mass matrix 
in the basis $(\nu_e, \nu_\mu, \nu_\tau)$ is of the form
\begin{equation}
{\cal M}_\nu = \left[ \begin{array} {c@{\quad}c@{\quad}c} a_1^2 & a_1 a_2 + b 
& a_1 a_3 \\ a_1 a_2 + b & a_2^2 & a_2 a_3 \\ a_1 a_3 & a_2 a_3 & a_3^2 
\end{array} \right],
\end{equation}
where we have assumed that the usual one-loop contributions from bilinear 
$R$-parity violation \cite{brpv} are actually negligible, which is the case 
for most of the MSSM parameter space.  This matrix has 4 parameters and 
yields 3 eigenvalues and 3 mixing angles. Consider for example $a_3 = a_2$ 
and define $x \equiv 1 + (b/a_1 a_2)$, we then have
\begin{equation}
{\cal M}_\nu = \left[ \begin{array} {c@{\quad}c@{\quad}c} a_1^2 & x a_1 a_2 & 
a_1 a_2 \\ x a_1 a_2 & a_2^2 & a_2^2 \\ a_1 a_2 & a_2^2 & a_2^2 \end{array} 
\right].
\end{equation}
Assuming that $a_1$ and $x a_1$ are much smaller than $a_2$, the eigenvalues 
are easily determined to be
\begin{eqnarray}
m_1 &=& - {(1-x) a_1 a_2 \over \sqrt 2} + {(1-x)(3+x) a_1^2 \over 8}, \\ 
m_2 &=& {(1-x) a_1 a_2 \over \sqrt 2} + {(1-x)(3+x) a_1^2 \over 8}, \\ 
m_3 &=& 2 a_2^2 + {(1+x)^2 a_1^2 \over 4},
\end{eqnarray}
corresponding to the eigenstates
\begin{eqnarray}
\nu_1 &=& {1 \over \sqrt 2} \left[ 1 - {(3+x) a_1 \over 8 \sqrt 2 a_2} \right] 
\nu_e + {1 \over 2} \left[ 1 - {(1+3x) a_1 \over 8 \sqrt 2 a_2} \right] 
\nu_\mu - {1 \over 2} \left[ 1 + {(7+5x) a_1 \over 8 \sqrt 2 a_2} \right] 
\nu_\tau , \\ 
\nu_2 &=& {1 \over \sqrt 2} \left[ 1 + {(3+x) a_1 \over 8 \sqrt 2 a_2} \right] 
\nu_e - {1 \over 2} \left[ 1 + {(1+3x) a_1 \over 8 \sqrt 2 a_2} \right] 
\nu_\mu + {1 \over 2} \left[ 1 - {(7+5x) a_1 \over 8 \sqrt 2 a_2} \right] 
\nu_\tau , \\ 
\nu_3 &=& {(1+x) a_1 \over 2 \sqrt 2 a_2} \nu_e + {1 \over \sqrt 2} \nu_\mu 
+ {1 \over \sqrt 2} \nu_\tau,
\end{eqnarray}
which is of course very near the case of bimaximal mixing.  Atmospheric 
neutrino oscillations are thus explained by $\nu_\mu \to \nu_\tau$ with 
$\sin^2 2 \theta \simeq 1$ and
\begin{equation}
\Delta m_{23}^2 \simeq \Delta m_{13}^2 \simeq 4 a_2^4 + {1 \over 2} 
(1 + 6x + x^2) a_1^2 a_2^2, 
\end{equation}
and solar neutrino oscillations by $\nu_e \to (\nu_\mu - \nu_\tau)/ 
\sqrt 2$ with $\sin^2 2 \theta \simeq 1$ and
\begin{equation} 
\Delta m_{12}^2 \simeq {(1-x)^2 (3+x) \over 2 \sqrt 2} a_1^3 a_2.
\end{equation}
Using $a_2 = 0.16$ eV$^{1/2}$, $a_1 = 0.05$ eV$^{1/2}$, and $x=-1$, we find 
$\Delta m_{atm}^2 \simeq 2.5 \times 10^{-3}$ eV$^2$, and $\Delta m_{sol}^2 
\simeq 5.7 \times 10^{-5}$ eV$^2$, in very good agreement with data.

Referring back to Fig.~2, we calculate the parameter $b$ to be given by
\begin{equation}
b = {G_F m_\mu^2 \over 4 \pi^2 \sqrt 2}  {h A m_\tau 
\langle \tilde \nu_\tau \rangle \over m^2_{eff} \cos^2 \beta},
\end{equation}
where $m_{eff}$ is a function of $m_{\tilde \tau^c}$ and $m_{h^\pm}$. 
Using $h = 2$ and $m_{eff}^2/A = 1$ TeV, we find that in order to  
obtain $b = -2 a_1 a_2 \simeq 0.016$ eV, we need $\langle \tilde \nu_\tau 
\rangle \simeq 1.93 \cos^2 \beta$ GeV.  This relatively small value is 
negligible compared to $v = (2 \sqrt 2 G_F)^{-1/2} = 174$ GeV (especially 
for large values of $\tan \beta$), and consistent with all present 
low-energy phenomenology.

The salient feature of our model is that $\tilde \nu_e$ must be relatively 
light, say around 200 GeV, to explain a large $\Delta a_\mu$.  In that 
case, $\tilde e$ must also be light, because of the well-known MSSM 
relationship
\begin{equation}
m^2_{\tilde e} = m^2_{\tilde \nu_e} - M_W^2 \cos 2 \beta.
\end{equation}
Now both $\tilde \nu_e$ and $\tilde e$ can be produced by electroweak 
interactions, such as $Z \to \tilde \nu_e^* \tilde \nu_e$ and $W^- \to 
\tilde \nu_e^* \tilde e$.  They must then decay according to Eq.~(2), i.e.
\begin{equation}
\tilde \nu_e \to \mu^+ \tau^-, ~~~ \tilde e \to \bar \nu_\mu \tau^-.
\end{equation}
These are very distinctive signatures and if observed, the two masses may 
be reconstructed and the value of $\beta$ determined by Eq.~(18).

If the MSSM neutralinos $\tilde \chi^0_i$ and charginos $\tilde \chi^+_i$ 
are produced, as decay products of squarks for example, then the decays
\begin{equation}
\tilde \chi^0_i \to \tilde \nu_e \bar \nu_e (\tilde \nu_e^* \nu_e), ~ 
\tilde e e^+ (\tilde e^* e^-), ~~~ \tilde \chi^+_i \to \tilde \nu_e e^+, ~ 
\tilde e^* \nu_e
\end{equation}
are possible.  The subsequent decays of Eq.~(19) would again be indicative 
of our model. In a future muon collider, the process
\begin{equation}
\mu^+ \mu^- \to \tilde \nu_e^* \tilde \nu_e 
\end{equation}
(through $\tau$ exchange) is predicted, by which the $\tilde \nu_e$ decay of 
Eq.~(19) could be studied with precision.

Single production of $\tilde \nu_e$ and $\tilde e$ is also possible in 
an $e^+ e^-$ collider.  There are 4 different final states:  $\tau^+ \mu^- 
\tilde \nu_e$, $\tau^+ \nu_\mu \tilde e$, and their conjugates.  With the 
subsequent decays given by Eq.~(19), the experimental signatures are 
4 charged leptons $(\tau^+ \tau^- \mu^+ \mu^-)$ and 2 charged taus + missing 
energy $(\tau^+ \tau^- \bar \nu_\mu \nu_\mu)$.  The absence of such events 
at LEP up to 207 GeV constrains $h$ and $m_{\tilde \nu_e}$.  Although a 
quantitative analysis is not available at present, we estimate the likely mass 
bound (on the basis that it would be similar to that of single scalar 
leptoquark production) to be around 180 GeV for $h=1$.  With such a large 
mass, we will need a larger $h$ to get $\Delta a_\mu \sim 10^{-9}$, so we 
have chosen $h=2$ and $m_{\tilde \nu_e} = 200$ GeV (which puts $\tilde \nu_e$ 
beyond the production capability of LEP) as representative values.

Lepton-flavor violating processes are very much suppressed in our model, 
because they have to be proportional to the small parameters $\mu_\alpha$ in 
Eq.~(6) or the small vacuum expectation values $\langle \tilde \nu_\alpha 
\rangle$.  For example, the rare decay $\tau \to e \gamma$ proceeds in 
one-loop order through $\tilde \nu_e$ exchange and the mixing of $\mu_L$ 
with $\tilde w^-$, i.e.
\begin{equation}
\left( {\mu_\mu \over \mu_0} - {\langle \tilde \nu_\mu \rangle \over 
v \cos \beta} \right) {M_W \cos \beta \sqrt 2 \over m_{\tilde w}},
\end{equation}
and through $\tilde e$ exchange and the mixing of $\nu_\mu$ with $\tilde B$ 
and $\tilde w^0$, i.e.
\begin{equation}
- \left( {\mu_\mu \over \mu_0} - {\langle \tilde \nu_\mu \rangle \over 
v \cos \beta} \right) {M_Z \sin \theta_W \cos \beta \over m_{\tilde B}}, 
~~~ \left( {\mu_\mu \over \mu_0} - {\langle \tilde \nu_\mu \rangle \over 
v \cos \beta} \right) {M_Z \cos \theta_W \cos \beta \over m_{\tilde w}},
\end{equation}
where $\mu_0$ is the coefficient of the $(\hat h_1^- \hat h_2^+ - \hat h_1^0 
\hat h_2^0)$ term in the superpotential of the MSSM.  
Its amplitude is approximately given by
\begin{equation}
{\cal A} =  { e h g (5 - \tan^2 \theta_W) \over 96 \pi^2 \sqrt 2} \left( 
{\mu_\mu \over \mu_0} - {\langle \tilde \nu_\mu \rangle \over v \cos \beta} 
\right) {M_W \cos \beta \over m_{eff}^2} \epsilon^\alpha q^\beta  \bar e 
\sigma_{\alpha \beta} \left( {1 + \gamma_5 \over 2} \right) \tau, 
\end{equation}
where
$m_{eff}$ is a function of all the heavy masses in the loop and normalized 
so that if all of them are equal, then they are all equal to $m_{eff}$. 

Since the neutrino mass parameter $a_2$ used earlier is given by
\begin{equation}
a_2^2 = \left( {\mu_\mu \over \mu_0} - {\langle \tilde \nu_\mu \rangle \over 
v \cos \beta} \right)^2 {M_Z^2 \cos^2 \beta \over m_{\tilde B} m_{\tilde w}} 
(m_{\tilde B} \cos^2 \theta_W + m_{\tilde w} \sin^2 \theta_W),
\end{equation}
the $\tau \to e \gamma$ branching fraction is related to it by
\begin{equation}
B(\tau \to e \gamma) = {\sin^2 \theta_W \cos^2 \theta_W \over 48 \pi^2} 
{(5 - \tan^2 \theta_W)^2 h^2 a_2^2 m_{\tilde B} m_{\tilde w} M_W^4 \over 
(m_{\tilde B} \cos^2 \theta_W + m_{\tilde w} \sin^2 \theta_W) m_\tau^2 
m_{eff}^4} ~B(\tau \to e \nu \bar \nu).
\end{equation}
Using $h=2$, $a_2 = 0.16$ eV$^{1/2}$, and assuming that $m_{\tilde B} = 
m_{\tilde w} = m_{eff} = 200$ GeV, we find $B(\tau \to e \gamma) \simeq 
2.5 \times 10^{-13}$, which is many orders of magnitude below the experimental 
upper bound of $2.7 \times 10^{-6}$.  

The $\mu \to e \gamma$ rate is even more suppressed because it has to 
violate both $L_\mu$ and $L_e$, whereas $\tau \to e \gamma$ only needs to 
violate $L_\mu$.  We note that if we had chosen the extra term in Eq.~(1) 
to be $h(\hat \nu_e \hat \tau - \hat e \hat \nu_\tau) \hat \mu^c$ or 
$h(\hat \nu_\mu \hat \tau - \hat \mu \hat \nu_\tau) \hat e^c$, then $\mu \to 
e \gamma$ would not be doubly suppressed and would have a branching fraction 
of about $4 \times 10^{-10}$, in contradiction with the present experimental 
bound \cite{meg} of $1.2 \times 10^{-11}$.  We note also that $m_{ee} = a_1^2$ 
of Eq.~(7) is the effective neutrino mass measured in neutrinoless double beta 
decay.  It is of order $10^{-3}$ eV in our model, which is well below the 
present experimental bound \cite{dbd} of 0.2 eV.

In conclusion, we have shown how a novel minimal extension of the MSSM with 
$L_\tau = L_e + L_\mu$ allows it to have a large contribution to the muon 
anomalous magnetic moment without otherwise constraining the usual MSSM 
parameter space.  With the soft and spontaneous breaking of this lepton 
symmetry, realistic neutrino masses (with bimaximal mixing) are generated 
for a natural explanation of atmospheric and solar neutrino oscillations. 
The scalar electron doublet $(\tilde \nu_e, \tilde e)$ is predicted to be 
light (perhaps around 200 GeV) and has distinctive experimental signatures.

This work was supported in part by the U.~S.~Department of Energy under 
Grant No.~DE-FG03-94ER40837.  R.A. acknowledges the support of D.S.T. India. 
G.R. also thanks the UCR Physics Department for hospitality and acknowledges 
INSA (New Delhi) for support.

\bibliographystyle{unsrt}

\begin{thebibliography}{99}
\bibitem{mn} E. Ma and D. Ng, Phys. Rev. {\bf D41} R1005 (1990).
\bibitem{g-2} H. N. Brown {\it et al.}, Phys. Rev. Lett. {\bf 86}, 2227 (2001).
\bibitem{susyg2} See for example A. Czarnecki and W. J. Marciano, Phys. Rev. 
{\bf D64}, 013014 (2001); S. P. Martin and J. D. Wells, Phys. Rev. {\bf D64} 
035003 (2001) and references therein.
\bibitem{atm} S. Fukuda {\it et al.}, Super-Kamiokande Collaboration, Phys. 
Rev. Lett. {\bf 85}, 3999 (2000) and references therein.
\bibitem{solar} S. Fukuda {\it et al.}, Super-Kamiokande Collaboration, Phys. 
Rev. Lett. {\bf 86}, 5656 (2001) and references therein.  See also Q. R. 
Ahmad {\it et al.}, SNO Collaboration, Phys. Rev. Lett. {\bf 87}, 071301 
(2001).
\bibitem{kkl} J. E. Kim, B. Kyae, and H. M. Lee, hep-ph/0103054; R. Adhikari 
and G. Rajasekaran, hep-ph/0107279.
\bibitem{bgh} V. Barger, G. F. Giudice, and T. Han, Phys. Rev. {\bf D40}, 
2987 (1989).
\bibitem{qcd} There are uncertainties regarding the quantum-chromodynamics 
contribution.  See for example J. Erler and M. Luo, Phys. Rev. Lett. {\bf 87}, 
071804 (2001); F. J. Yndurain, hep-ph/0102312; S. Narison, Phys. Lett. 
{\bf B513}, 53 (2001); F. Jegerlehner, hep-ph/0104304; K. Melnikov, 
hep-ph/0105267.
\bibitem{cf} See for example C.-H. Chang and T.-F. Feng, Eur. Phys. J. 
{\bf C12}, 137 (2000) and references therein.
\bibitem{drv} M. A. Diaz, J. C. Romao, and J. W. F. Valle, Nucl. Phys. 
{\bf B524}, 23 (1998).
\bibitem{brpv} D. E. Kaplan and A. E. Nelson, JHEP {\bf 0001}, 033 (2000); 
F. Takayama and M. Yamaguchi, Phys. Lett. {\bf B476}, 116 (2000); M. Hirsch, 
M. A. Diaz, W. Porod, J. C. Romao, and J. W. F. Valle, Phys. Rev. {\bf D62}, 
113008 (2000); O. C. W. Kong, JHEP {\bf 0009}, 037 (2000).
\bibitem{meg} M. L. Brooks {\it et al.}, Phys. Rev. Lett. {\bf 83}, 1521 
(1999).
\bibitem{dbd} L. Baudis {\it et al.}, Phys. Rev. Lett. {\bf 83}, 41 (1999).

\end{thebibliography}


\newpage
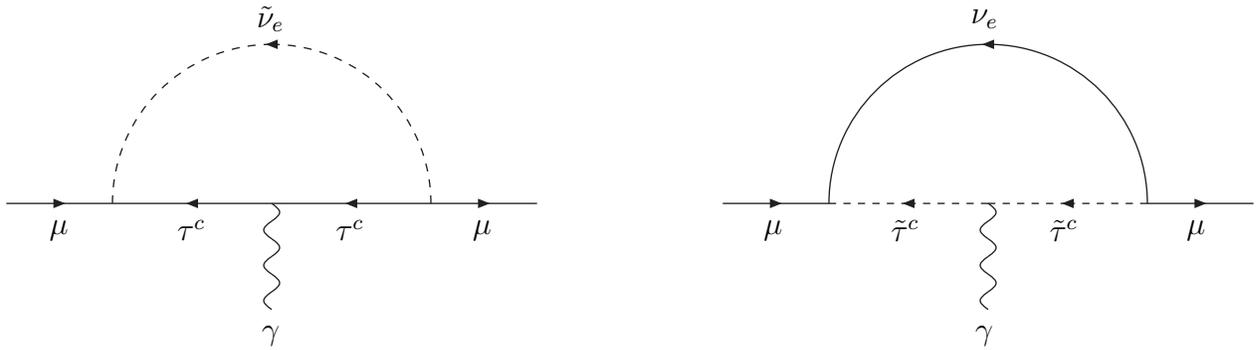
\begin{figure}
\begin{center}
\begin{picture} (600,150) (0,0)
\ArrowLine(5,40)(45,40)
\ArrowLine(105,40)(45,40)
\ArrowLine(165,40)(105,40)
\ArrowLine(165,40)(205,40)
\Photon(105,40)(105,0){3}{3}
\DashArrowArc(105,40)(60,0,180){3}
\Text(25,30)[]{$\mu$}
\Text(75,30)[]{$\tau^c$}
\Text(135,30)[]{$\tau^c$}
\Text(185,30)[]{$\mu$}
\Text(105,-10)[]{$\gamma$}
\Text(105,110)[]{$\tilde \nu_e$}

\ArrowLine(275,40)(315,40)
\DashArrowLine(375,40)(315,40){3}
\DashArrowLine(435,40)(375,40){3}
\ArrowLine(435,40)(475,40)
\Photon(375,40)(375,0){3}{3}
\ArrowArc(375,40)(60,0,180)
\Text(295,30)[]{$\mu$}
\Text(345,30)[]{$\tilde \tau^c$}
\Text(405,30)[]{$\tilde \tau^c$}
\Text(455,30)[]{$\mu$}
\Text(375,-10)[]{$\gamma$}
\Text(375,110)[]{$\nu_e$}

\end{picture}
\end{center}
\caption{Contributions to muon anomalous magnetic moment.}
\end{figure}

\begin{figure}
\begin{center}
\begin{picture}(300,150)(0,0)
\ArrowLine(45,40)(85,40)
\DashArrowLine(145,40)(85,40){3}
\DashArrowLine(145,40)(205,40){3}
\ArrowLine(245,40)(205,40)
\DashArrowLine(145,40)(145,0){3}
\ArrowArc(145,40)(60,90,180)
\ArrowArcn(145,40)(60,90,0)
\Text(65,30)[]{$\nu_e$}
\Text(115,30)[]{$\tilde \tau^c$}
\Text(175,30)[]{$h_1^-$}
\Text(225,30)[]{$\nu_\mu$}
\Text(145,100)[]{$\times$}
\Text(95,90)[]{$\mu$}
\Text(195,90)[]{$\mu^c$}
\Text(145,-10)[]{$\langle \tilde \nu_\tau \rangle$}
\end{picture}
\end{center}
\caption{Radiative contribution to the $\nu_e \nu_\mu$ mass.}
\end{figure}
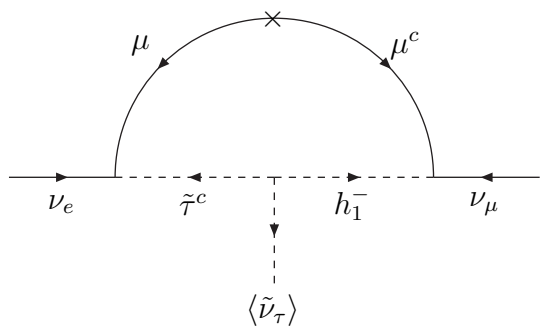

\end{document}